\documentclass[12pt]{article} 
\usepackage[tablesfirst,notablist,nomarkers]{endfloat} 

\usepackage{ol2}
\usepackage[draft]{hyperref}
\usepackage{amsmath}

\begin{document}


\title{Tunable photonic Bloch oscillations in electrically modulated photonic crystals}


\author{Gang Wang,$^{1,2}$ Ji Ping Huang,$^{1,2,*}$ and Kin Wah Yu$^{1,**}$}

\address{
$^1$Department of Physics, Chinese
University of Hong Kong, Shatin, New Territories, Hong Kong
\\
$^2$Surface Physics Laboratory  and Department of Physics, Fudan
University, Shanghai 200433, China \\
$^*$Corresponding author: jphuang@fudan.edu.cn\\
$^{**}$Corresponding anthor: kwyu@phy.cuhk.edu.hk
}

\begin{abstract}

We exploit theoretically the occurrence and tunability of photonic
Bloch oscillations (PBOs) in one-dimensional photonic crystals
(PCs) containing nonlinear composites. Because of the enhanced
third-order nonlinearity (Kerr type nonlinearity) of composites,
photons undergo oscillations inside tilted  photonic bands, which
are achieved by the application of graded external pump electric
fields on such PCs, varying along the direction perpendicular to
the surface of layers. The tunability of PBOs (including amplitude
and period) is readily achieved by changing the field gradient.
With an appropriate graded pump AC or DC electric field, terahertz
PBOs can appear and cover a terahertz band in electromagnetic
spectrum.

\end{abstract}

\ocis{230.4910, 350.4238, 350.5500.}


\noindent Recently matter wave dynamics of Bloch oscillations has
already motivated a good deal of attention. In 1928, Bloch made a
striking prediction that in a crystal lattice, a homogeneous static
electric field induces an oscillatory rather than uniform motion of
the electrons~\cite{Bloch}, referred to as Bloch oscillations (BOs).
Electronic BOs have been observed in semiconductor
superlattices~\cite{Esaki,Feldmann}. Now this phenomenon has been
extended to various classical wave systems, such as acoustic
systems~\cite{AlepuzPRL2007}, elastic
systems~\cite{MoralesPRL2006}, and photonic
systems~\cite{SapienzaPRL2003}. The
electronic BOs are quite easily tunable via externally applied
electric fields or magnetic fields, or simultaneously by
both~\cite{Grill2007}. However, for uncharged particles like
photons, one has to resort to other approaches for
controllability~\cite{FanPRB2005}.

In this work, we propose a  class of PCs to realize the occurrence
and tunability of photonic Bloch oscillations (PBOs) by graded pump
DC or AC electric fields ${\bf E}_0$. The idea relies on the fact
that the effective dielectric permittivity (thus refractive index)
of some materials depends on external electric fields because of the
third-order nonlinearity (Kerr type nonlinearity). To exemplify this idea,
we theoretically calculate the PBOs in such PCs under different
graded pump electric fields. Compared with existing
proposals~\cite{FanPRB2005}, our proposal offers ultrafast response
time and dynamical control. Also, we show that with an
appropriate choice of parameters, this could lead to a generation of
teraherz radiations.

Figure 1 shows a one-dimensional double-layer PC, which consists of
alternative layers of composite materials~\cite{Wang} (or composites
for short, e.g., Au/SiO2) and common dielectric (e.g., air). Through
local field and resonant scattering effects in
nanoparticles~\cite{HuangPhysRep2006}, the third-order
nonlinearity  of composites can be
enhanced~\cite{RichardOptlett1985,HuangPhysRep2006}.

\begin{figure}[htb]
\centerline{\includegraphics[width=8cm]{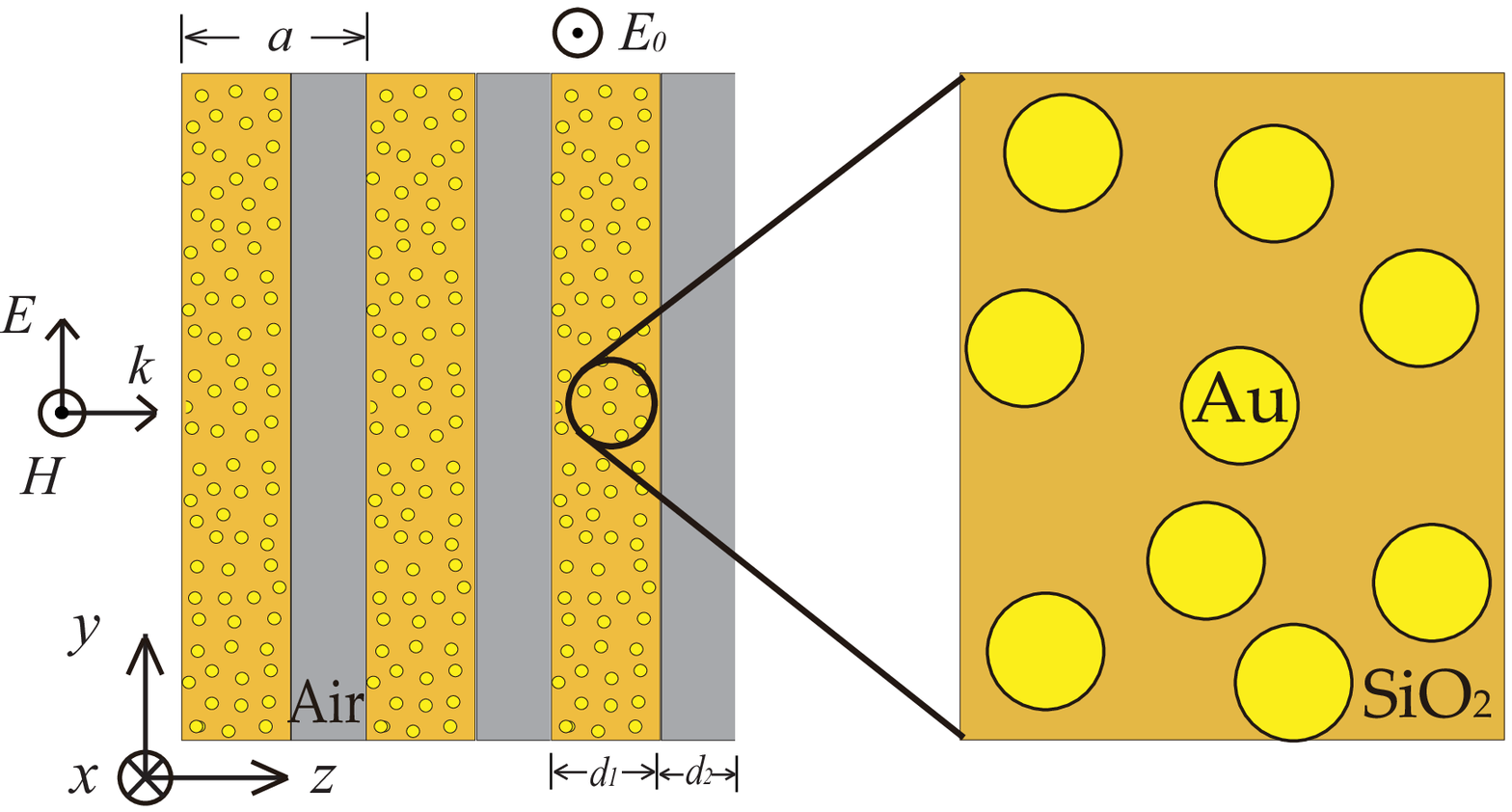}} \caption{(Color
online) Schematic view of the one-dimensional PC composed of  a
composite layer (nonlinear composites) with thickness $d_1$ and a
dielectric layer (e.g., air) with thickness $d_2$. The composites
can be prepared with nonlinear nanoparticles (e.g., gold)
randomly dispersed in a linear dielectric (e.g., silica), as
shown in the right panel. For plotting
Figs.~\ref{fig2}-\ref{fig4}, we shall use $d_1=d_2=0.5a$, where
$a$ denotes the lattice constant.}\label{fig1}
\end{figure}

Let us denote by ${\bf D}_{0 (1)}$ the response to pump (probe)
field ${\bf E}_{0 (1)}$. For weak nonlinearity under consideration,
namely $\chi^{(3)}{\bf |E|}^2\ll\epsilon$, the constitutive relation
${\bf D}=\epsilon {\bf E} + \chi^{(3)} {\bf |E|}^2 {\bf E}$ can be
written as ${\bf D}={\bf D}_0+\alpha {\bf D}_1$, and ${\bf E}={\bf
E}_0+\alpha {\bf E}_1$, where $\alpha$ is a small parameter. Due to
the weak nonlinearity limit, we apply the perturbation
method~\cite{YuPRB1992} to express ${\bf D}_0$ and ${\bf D}_1$ in
terms of ${\bf E}_0$ and ${\bf E}_1$: ${\bf
D}_0=(\epsilon+\chi^{(3)} |{\bf E}_0|^2){\bf E}_0$, ${\bf
D}_1=\epsilon {\bf E}_1+\chi^{(3)} |{\bf E}_0|^2 {\bf
E}_1+2\chi^{(3)} \rm{Re[E_0^*E_1]}\cos\theta{\bf E}_0$. Since the
pump field ${\bf E}_0$ is much larger than the probe field ${\bf
E}_1$, the  response to the probe field ${\bf D}_1$ is rather
stable, and it is related to the angle $\theta$ between the pump
field (${\bf E}_0$) and the probe field (${\bf E}_1$). Without loss
of generality, we assume ${\bf E}_1\| {\bf E}_0$ (see
Fig.~\ref{fig1}), which results in an effective dielectric constant
for the probe field
$\tilde{\epsilon}_{eff}=\epsilon+3\chi^{(3)}|{\bf E}_0|^2$. So the
dielectric constant  $\tilde{\epsilon}_p$ possessed by nonlinear
nanoparticles can be expressed as
\begin{equation}
\tilde{\epsilon}_p=\epsilon_{p}+3\chi_p^{(3)}|\textbf{E}_p|^2
\approx
\epsilon_{p}+3\chi_p^{(3)}\langle|\textbf{E}_p|^2\rangle,\label{eq1}
\end{equation}
where $\epsilon_{p}$ denotes the linear (field-independent)
dielectric permittivity, $\chi_p^{(3)}$ the third-order nonlinear
susceptibility of the nanoparticles, ${\bf E}_p$ the local electric
field inside the nanoparticle, and $\langle\cdots\rangle$ the volume
average of $\cdots$.  Then, the effective nonlinear permittivity
$\tilde{\epsilon}_{1}$ of the layer of composites can be given by
the Maxwell-Garnett
approximation~\cite{HuangPhysRep2006}
\begin{equation}
\frac{\tilde{\epsilon}_{1}-\epsilon_{h}}{\tilde{\epsilon}_{1}+2\epsilon_{h}}=p
\frac{\tilde{\epsilon}_{p}-\epsilon_{h}}{\tilde{\epsilon}_{p}+2\epsilon_{h}},
\end{equation}
where  $\epsilon_{h}$ represents the dielectric permittivity of the
host (which is assumed to be linear for simplicity), and $p$ the
volume fraction of nanoparticles. The volume-averaged local electric
field $\langle|{\bf E}_p|^2\rangle$ in the layer is given
by~\cite{ShengPingJOSAB1998,YuSSC98}
\begin{equation}
\langle|{\bf E}_p|^2\rangle =
\frac{9|\epsilon_{h}|^2}{|(1-p)\epsilon_p+(2+p)\epsilon_{h}|^2}{\bf
E}_0^2\equiv \beta^2 {\bf E}_0^2.\label{equation}
\end{equation}
Thus the effective nonlinear response  of the layer can be much
enhanced. 

For model calculations, we investigate the composite of Au/SiO$_2$
with volume fraction of Au nanoparticles $p=0.20$. The dielectric
permittivities of SiO$_2$ and air (dielectric layer) are taken to be
$\epsilon_h = 2.25\epsilon_0$ and $\epsilon_{air} = \epsilon_0$. The
probe  field can be much weaker in strength than the pump electric
field, so that the dispersion (and loss) of the (much weaker) probe
field can be neglected. Here we take
$\epsilon_p=-9.97\epsilon_0$~\cite{ShengPingAPL1997,ShengPingJOSAB1998},
which is a real frequency-independent constant. This corresponds to
a pump field produced by a laser source at
$620\,$nm~\cite{ShengPingJOSAB1998}.

\begin{figure}[htb]
\centerline{\includegraphics[width=6cm]{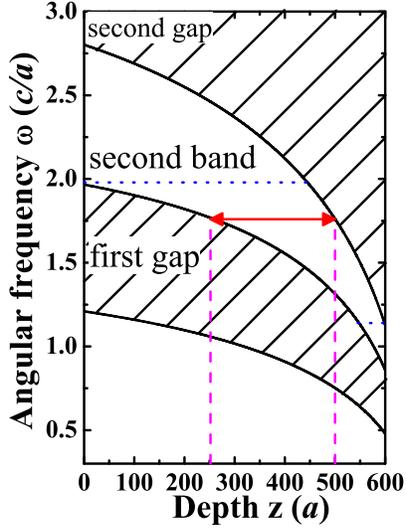}} \caption{(Color
online) Depth-dependent photonic band structure for gradation
profile $\chi^{(3)}_p E_0^2=g z+b$ with parameters
$g=3.67\times10^{-4}$ and $b=1.33\times10^{-2}$ [which correspond to
Fig.~\ref{fig4}(a)]. The vertical dashed lines label the space of
oscillations displayed in Fig.~\ref{fig4}(a), for an incident wave
with $\omega_0=1.76$ $c/a$. The cross section composed by the two
horizontal dotted lines in the second band denotes the full region
for the appearance of oscillations for various angular frequencies
of incident electromagnetic waves.}\label{fig2}
\end{figure}

\begin{figure}[htb]
\centerline{\includegraphics[width=8cm]{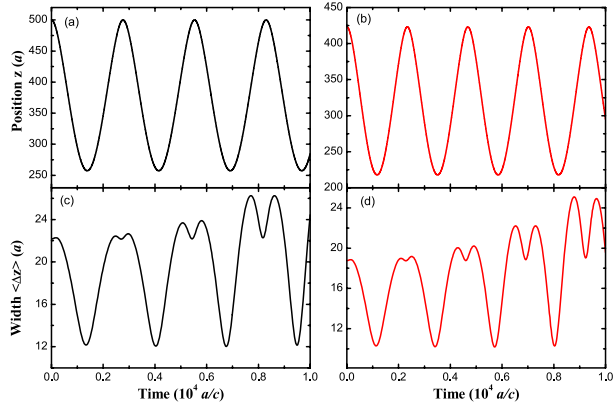}} \caption{(Color
online) The time-dependent mean position and width of a wave
packet under pump electric fields with different gradation
profiles $\chi^{(3)}_p E_0^2=gx+b$. The panels (a) and (c)
correspond to $g=3.67\times10^{-4}$ and $b=1.33\times10^{-2}$,
(b) and (d) for $g=4.33\times10^{-4}$ and $b=1.33\times10^{-2}$.
The Bloch oscillations clearly develop around different centers
with different periods.}\label{fig4}
\end{figure}

The existence of the enhanced Kerr  nonlinearity  enables us to
modify photonic band structures of such PCs by applying pump
electric fields~\cite{Wang}. Throughout this work, we use
$\chi_p^{(3)}E_0^2$ to indicate the strength of external pump
electric field $E_0$. To obtain PBOs, we apply an external electric
field with gradation profile $\chi_p^{(3)}E_0^2=gz+b$, which can be
adjusted to achieve different gradation profiles. In graded fields,
the photonic band structures become depth-dependent, viz.,
$z$-dependent (see Fig.~\ref{fig2}), which can be evaluated via the
transfer matrix method with $g=3.67\times10^{-4}$ and
$b=1.33\times10^{-2}$. The reason is easily understood. When the
pump electric field acting on the layers of composites varies along
the $z$ direction, the effective dielectric permittivity of each
composite layer grows monotonically, resulting in a refractive index
gradient. By carefully examining the condition for the appearance of
PBOs, we are convinced that PBOs cannot happen in the lowest band.
Thus we will study PBOs in the second (or higher) band. Besides the
tilted band structures in Fig.~\ref{fig2}, the other prerequisite
for the appearance of PBOs is that $\omega(z_{max},~k=0)
=\omega(z_{min},~k=\pi/a)$, where $\omega$ represents the angular
frequency of source waves, $z_{max(min)}$ means the  maximum
(minimum) position for PBOs, and $k$ is the Bloch wave vector. In
other words, PBOs just appear within the cross section that is
composed by the two horizontal dotted lines as displayed in
Fig.~\ref{fig2}. So when the incidence with $\omega_0=1.76$ $c/a$
illuminates the structure (see the arrow in Fig~\ref{fig2}), the
oscillations occure in the spatial range $250a< z< 500a$, in good
agreement with the numerical results shown in Fig.~\ref{fig4} (a).

The above analysis shows that PBOs can appear inside the second band
if a graded external electric field is applied on such 1D PCs. When
electromagnetic waves are incident on the PCs, multiple reflections
on the gaps lead to spatial BOs. The actual calculation is to solve
for the propagation of a Gaussian wavepacket initially peaked around
$z=z_{max}$ and $k=0$ at $t=0$. The subsequent motion of the
wavepacket is governed by the Hamiltonian equations of motion of
$\omega(z,k)$. At $t=T_B/2$ ($T_B$: time period of PBOs), $z$
becomes $z_{min}$ while $k$ reaches $\pi/a$ and PBO occurs. The
Hamiltonian equations of motion of $\omega(z,k)$ have been
integrated with appropriate initial conditions to yield the time
series of displacement $z(t)$ and momentum $k(t)$ (not shown
herein). As expected, such oscillations are clearly seen from the
mean position $\langle z(t) \rangle$ in Fig.~\ref{fig4} (a) and (b).
For the use and justification of the Hamiltonian formalism for
optics, please see Ref.~\cite{Russell} and references therein. Based
on the semiclassical solutions, we access the width $\Delta z(t)$ of
a wave packet and show that it remains bound in time. For a packet
with Gaussian distribution with widths $\sigma_z = 5$ and $\sigma_k
= 0.2$, we show the time dependent mean width $\Delta z(t)$
Fig.~\ref{fig4} (c) and (d). We can find that the width is about
5$\%$ of the length of the superlattice (600$a$) and increases only
slightly after several periods, which is acceptable. This originates
from the non-constant inclination of the band diagram. Thus we can
be convinced that Bloch oscillation indeed occurs. In fact, we can
design linearly tilted bands to avoid the increase of the width.

The enhancement of the third-order nonlinearity of the composites
enables us to tune PBOs by using tunable inclined band structures.
Figure~\ref{fig4} shows the spatial range of PBOs for a fixed
incident frequency $\omega_0=1.76$ $c/a$  under different gradation
profiles. We find that the variation of gradient plays an important
role in the occurrence of PBOs, including their amplitude and
oscillation frequency. A key parameter for PBOs is oscillation
period $T_B$. Figure~\ref{fig4} also shows that $T_B$ depends on the
tuning parameters in the gradation profiles. Apparently, $T_B$
decreases while increasing the field gradient because of steeper
tilting bands. Therefore, there is a critical gradient
$g=2.73\times10^{-4}$, below which $T_B$ becomes infinite (or,
alternatively, no oscillations come to appear).

Considering the large nonlinear susceptibility (typical value $8
\times 10^{-8}$ esu) of Au nanoparticles~\cite{ShengPingJOSAB1998},
for the two gradation profiles adopted in our manuscript
$|E|^2=(gz+b)/\chi^{(3)}$, the maximum pump fields intensities are
$I_m=|E|^2= 61.5$ and $72.1$ mW/cm$^2$ respectively. They are
experimentally practical. The intensity of probe fields
can be fraction of milliwatts per cm$^2$, which is the requirement
of the weak nonlinearity. When the probe fields are at 535 nm, and
the thickness of each layer of the superlattice $d_1=d_2=75\,$nm,
the BO frequencies $f_B = 1/T_B$ are 0.720\,THz and 0.853\,THz
respectively. All the $f_B$'s are just located within the range of
terahertz band, namely, $10^{11} -10^{13}\,$Hz. For decoupling THz
radiation from the structure, one can apply a uniform pump field to
the structure, leading to the non-tilted band diagram. Meanwhile,
the structure is illuminated by an incident light with frequency in
the region of photonic allowed bands. Second, once the graded pump
electric field is applied suddenly, the pulse propagating in the
structure will be trapped and the oscillations commence
subsequently. Third, after several oscillation periods, the graded
pump field is restored to the uniform case, then the carrier waves
with terahertz modulation can escape from the structure. Meanwhile,
to avoid the leakage of energy in the transverse direction, spatial
confinement is needed.  In this regard, total internal reflection
like in graded-index optical fibers can be used. For low-index
medium, it is useful to guide light by means of a photonic band gap.


In summary, we have  theoretically exploited the occurrence and
tunability of PBOs in
one-dimensional PCs containing nonlinear
composites by  applying graded external electric fields.
Meanwhile, a kind of terahertz PBOs can appear and cover a
terahertz band in electromagnetic spectrum.

{\it Acknowledgments.}  This work was supported by the RGC Earmarked
Grant of Hong Kong SAR Government, by the C. N. Yang Fellowship in
CUHK, by the CNKBRSF under Grant No. 2006CB921706, by the Pujiang
Talent Project (No. 06PJ14006) of the Shanghai Science and
Technology Committee,  by the Shanghai Education Committee ("Shu
Guang" project), and by the NNSFC under Grant No. 10604014.

\end{document}